\documentclass[acmsmall,screen,authorversion,nonacm]{acmart} 
\usepackage{microtype}
\usepackage{xspace}
\usepackage{graphicx}
\graphicspath{{figures/}}
\usepackage{needspace}
\newcommand{\needlines}[1]{\Needspace{#1\baselineskip}}
\usepackage{paralist}
\usepackage{ifthen}
\usepackage[normalem]{ulem} 
\usepackage{xcolor}

\newboolean{showedits}
\setboolean{showedits}{true} 
\ifthenelse{\boolean{showedits}}
{
	\newcommand{\del}[1]{\textcolor{red}{\sout{#1}}} 
	\newcommand{\nbe}[3]{
		{\colorbox{#3}{\bfseries\sffamily\scriptsize\textcolor{white}{#1}}}
		{\textcolor{#3}{\sf\small$\blacktriangleright$\textit{#2}$\blacktriangleleft$}}}
}{
	\newcommand{\del}[1]{} 
	
	\newcommand{\nbe}[3]{}
}
\usepackage[most]{tcolorbox}
\ifthenelse{\boolean{showedits}}
{
  \newtcolorbox{inserted}{%
       title=Inserted text:,
       colframe=blue,colback=blue!5!white,
       breakable,
       leftrule=0mm, 
       bottomrule=0mm,
       rightrule=0mm,
       toprule=0mm,
       arc=0mm, outer arc=0mm,
       oversize
  }
  \newtcolorbox{deleted}{%
       title=Deleted text:,
       colframe=red,colback=red!5!white,
       breakable,
       leftrule=0mm, 
       bottomrule=0mm,
       rightrule=0mm,
       toprule=0mm,
       arc=0mm, outer arc=0mm,
       oversize
  }
  \newtcolorbox{refactored}{%
       title=Rewritten text:,
       colframe=blue,colback=red!5!white,
       breakable,
       leftrule=0mm, 
       bottomrule=0mm,
       rightrule=0mm,
       toprule=0mm,
       arc=0mm, outer arc=0mm,
       oversize
  }
}{

}
\newboolean{showcomments}
\setboolean{showcomments}{true}
\newcommand{\id}[1]{$-$Id: scgPaper.tex 32478 2010-04-29 09:11:32Z oscar $-$}

\ifthenelse{\boolean{showcomments}}
{\newcommand{\nbc}[3]{
 {\colorbox{#3}{\bfseries\sffamily\scriptsize\textcolor{white}{#1}}}
 {\textcolor{#3}{\sf\small$\blacktriangleright$\textit{#2}$\blacktriangleleft$}}}
 }
{\newcommand{\nbc}[3]{}
 }

\newboolean{isblinded}
\setboolean{isblinded}{true}
\ifthenelse{\boolean{isblinded}}
{\newcommand\blind[1]{BLINDED\xspace}}
{\newcommand\blind[1]{#1\xspace}}

\newcommand{\ie}{\emph{i.e.},\xspace}

\newcommand{\etal}{\emph{et al.}\xspace}

\newcommand{\sep}{\mbox{$\gg$}}
\usepackage[english]{babel}
\usepackage{xcolor}
\usepackage{listings}
\lstdefinelanguage{Smalltalk}{
  morestring=[d]',
  morecomment=[s]{"}{"},
  alsoletter={\#:},
  escapechar={!},
  literate=
    {BANG}{!}1
    {UNDERSCORE}{\_}1
    {_}{{$\leftarrow$}}1
    {>>>}{{\sep}}1
    {^}{{$\uparrow$}}1
    {~}{{$\sim$}}1
    {-}{{\sf -\hspace{-0.13em}-}}1  
    {+}{\raisebox{0.08ex}{+}}1		
    {-->}{{\quad$\longrightarrow$\quad}}3
	, 
  tabsize=4
}[keywords,comments,strings]

\definecolor{source}{gray}{0.95}

\newcommand{\st}{\lstinline[mathescape=false,backgroundcolor=\color{white},basicstyle={\sffamily\upshape}]}
\newcommand{\lst}[1]{{\textsf{\textup{#1}}}}
\lstnewenvironment{code}{%
	\lstset{%
		frame=single,
		framerule=0pt,
		mathescape=false
	}
}{}

\newboolean{preprint}
\setboolean{preprint}{true}
\ifthenelse{\boolean{preprint}}{
}{
}
\AtBeginDocument{%
  }
\ifthenelse{\boolean{preprint}}{
\setcopyright{acmlicensed}
}{
\setcopyright{ACMUNKNOWN} 
}
\acmDOI{10.1145/3689492.3690044}
\acmYear{2024}
\copyrightyear{2024}
\acmISBN{979-8-4007-1215-9/24/10}
\acmConference[Onward! '24]{Proceedings of the 2024 ACM SIGPLAN International Symposium on New Ideas, New Paradigms, and Reflections on Programming and Software}{October 23--25, 2024}{Pasadena, CA, USA}
\acmBooktitle{Proceedings of the 2024 ACM SIGPLAN International Symposium on New Ideas, New Paradigms, and Reflections on Programming and Software (Onward! '24), October 23--25, 2024, Pasadena, CA, USA}
\acmSubmissionID{onward24papers-p5-p}
\received{2024-04-25}
\received[accepted]{2024-08-08}
\newcommand*{\smallimg}[1]{%
    \raisebox{-.3\baselineskip}{%
        \includegraphics[
        height=\baselineskip,
        width=\baselineskip,
        keepaspectratio,
        ]{#1}%
    }%
}

\usepackage{caption}
\newcommand{\GT}{\lst{GT}\xspace} 
\newcommand\lmaf{\lst{Ludo\-Move\-Assert\-ion\-Fail\-ure}\xspace}
\newboolean{anonymous}
\setboolean{anonymous}{true}
\newcommand\anonymize[2]{\ifthenelse{\boolean{anonymous}}{#2}{#1}\xspace}

\newcommand\deet{{\tt deet}\xspace}
\begin{document}
\ifthenelse{\boolean{preprint}}{
\title[Moldable Exceptions --- preprint]{Moldable Exceptions}\thanks{Preprint of paper presented at Onward!, Pasadena, CA, Oct.\ 20-25, 2024.}
}{
\title{Moldable Exceptions}
}

\author{Andrei Chi\c{s}}
\affiliation{%
  \institution{feenk gmbh}
  \city{Wabern}
  \country{Switzerland}}
\email{andrei.chis@feenk.com}
\author{Tudor G\^irba}
\affiliation{%
  \institution{feenk gmbh}
  \city{Wabern}
  \country{Switzerland}}
\email{tudor.girba@feenk.com}
\author{Oscar Nierstrasz}
\affiliation{%
  \institution{feenk gmbh}
  \city{Wabern}
  \country{Switzerland}}
\email{oscar.nierstrasz@feenk.com}

\begin{abstract}
Debugging is hard.
Interactive debuggers are mostly the same.
They show you a stack, a way to sample the state of the stack, and, if the debugger is live, a way to step through execution.
The standard interactive debugger for a general-purpose programming language provided by a mainstream IDE mostly offers a low-level interface in terms of generic language constructs to track down and fix bugs.
A custom debugger, such as those developed for specific application domains, offers alternative interfaces more suitable to the specific execution context of the program being debugged.
Custom debuggers offering contextual debugging views and actions can greatly improve our ability to reason about the current problem.
Implementing such custom debuggers, however, is non-trivial, and poses a barrier to improving the debugging experience.
In this paper we introduce \emph{moldable exceptions}, a lightweight mechanism to adapt a debugger's interface based on contextual information provided by a raised exception.
We present, through a series of examples, how moldable exceptions can enhance a live programming environment.
\end{abstract}




\maketitle

\section{Introduction}\label{sec:intro}





Debuggers are unloved beasts.
They provide a generic and low-level interface to explore and debug the run-time state of a running program.
Many debugging problems can be better tackled by understanding the nature of the exception that was raised and caused the debugger to be activated.

There have been numerous efforts to develop custom debuggers for various application domains and domain-specific languages.
These custom debuggers provide dedicated views and actions that are tailored to a specific application context.
For example, an \emph{object-centric debugger}~\cite{Ress12a} offers views and interactions that focus on specific objects rather than the run-time stack.
Building a custom debugger is, however, a non-trivial task, so this does not happen too often.
An \emph{extensible} debugger (such as \deet~\cite{Hans97a}) is designed so that it can be easily extended with new graphical views and debugging operations, but these extensions still represent a significant development effort.
A \emph{moldable debugger}~\cite{Chis15c} is a special kind of extensible debugger that can activate alternative debugger interfaces depending on the current execution context, however the development of these alternative debuggers is still non-trivial.

We propose a new, lightweight mechanism, called \emph{moldable exceptions}, to dynamically adapt a moldable debugger using contextual information provided by the exception itself.
In modern, object-oriented software, it is common practice to define dedicated classes of exceptions to signal individual run-time issues.
Each exception therefore implicitly carries knowledge about the kind of issue being raised.
For example in a parsing framework an exception can indicate that a given input cannot be parsed.
Moldable exceptions leverage this knowledge by associating simple views and actions to be activated by a moldable debugger when that exception is raised.

\begin{figure}[h]
  \includegraphics[width=\columnwidth]{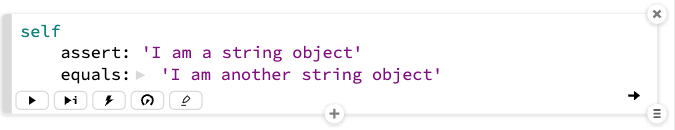}
  \caption{A failing string comparison assertion.}
  \label{fig:stringComparisonSnippet}
\end{figure}

\begin{figure}[h]
  \includegraphics[width=\columnwidth]{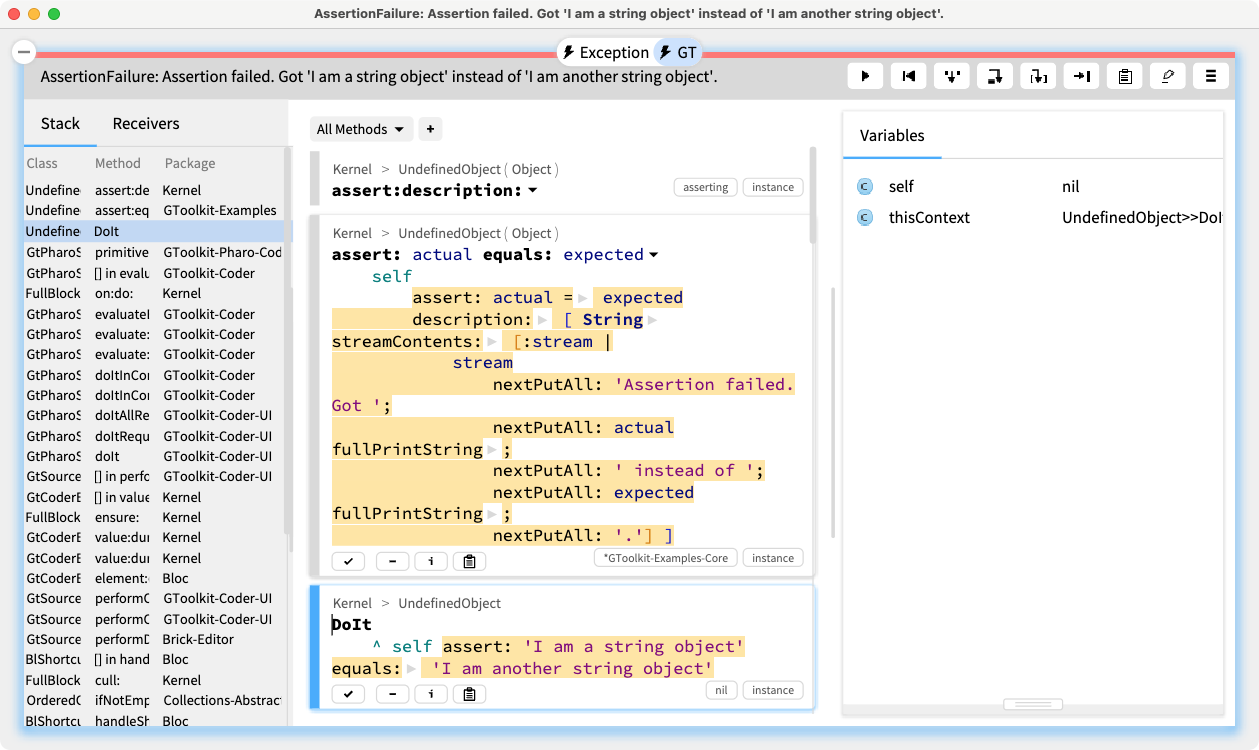}
  \caption{A generic debugger view.}
  \label{fig:genericDebugger}
\end{figure}

Consider the following example.
In \autoref{fig:stringComparisonSnippet} we see an assertion that compares two strings.\footnote{All the examples are written in Pharo Smalltalk (\url{https://pharo.org}), running in the open-source Glamorous Toolkit (\GT) IDE (\url{https://gtoolkit.com}).}
In a normal setting, this assertion will fail, yielding a standard debugger view, such as the one we see in \autoref{fig:genericDebugger}.
This is the standard debugger of the \GT IDE, which resembles other generic debuggers in that it shows an error message at the top, a compact view of the reified run-time stack at the left, source code of each stack frame in the middle, an inspector on the variables of the current stack frame at the right, and a dashboard of buttons at the top right for stepping through the running program.
It will typically take a developer some time to putter around in the debugger interface to understand the specific error (the strings do not match), and \emph{why} the strings do not match.

\begin{figure}[h]
  \includegraphics[width=\columnwidth]{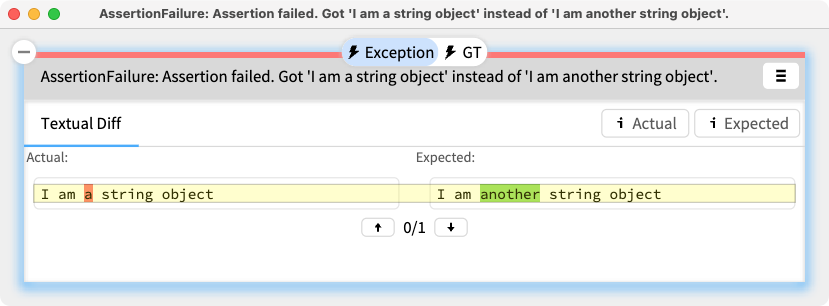}
  \caption{A string diff debugger view.}
  \label{fig:stringComparisonView}
\end{figure}

Suppose that instead of seeing the generic debugger, we are offered a view that highlights the actual differences, as in \autoref{fig:stringComparisonView}.
Such a view not only homes directly in on the specific problem, but also highlights the individual differences in a dedicated ``diff'' view.
Furthermore, since the diff view already exists as a component used in other applications, the development effort is close to zero.
While in this introductory example, the difference is easy to spot, in other cases such as those presented in \autoref{sec:casestudy}, a diff view is essential to quickly identifying differences.

Moldable exceptions work as follows: when an exception is raised, the exception (an object) is caught and passed to the debugger.
Every exception not only provides the debugger with the context it needs to generate the debugging UI, but it can also offer alternative views and actions.
In the case of our implementation, this is achieved by the exception class providing specially annotated debugger extension methods.

This simple mechanism allows moldable exceptions to do three things:
\begin{enumerate}[(i)]
	\item provide domain-specific debugging views and actions,
	\item offer new debugger GUI interactions, and
	\item enable automated fixes (code and data transformations) for common programming errors.
\end{enumerate}

We will illustrate these three points with several examples.
In \autoref{sec:inspectorViews} we introduce the \emph{moldable inspector}, which allows objects to define custom views and actions when they are inspected.
In \autoref{sec:views} we show how custom debugger views and actions can easily be defined in a similar way by adding simple annotated view and action methods to the class of a moldable exception.
In \autoref{sec:interactions} we show how a richer debugger user interface can be provided in much the same way by leveraging existing GUI frameworks.
We show an example of moldable exceptions enabling automated fixes in \autoref{sec:fixes}.
In \autoref{sec:casestudy} we describe a case study in which we used moldable exceptions to improve a development workflow.
We discuss the implementation details in \autoref{sec:implementation}.
We summarize our contributions and discuss some possible future work in \autoref{sec:directions}, and we discuss related work in \autoref{sec:related}.
We conclude in \autoref{sec:conclusion}.

\section{The Moldable Inspector}\label{sec:inspectorViews}

Understanding moldable exceptions can be easier if we first introduce another example of a similar approach.
Moldable exceptions build on the idea of a \emph{moldable tool}~\cite{Chis17a}, an IDE tool that adapts its behavior to a specific run-time application context.
An example of such a tool is a \emph{moldable object inspector}~\cite{Chis15a}.
When a moldable object inspector is opened on an object it looks for any view or action methods defined in the class of that object, and uses those methods to create custom views.

\begin{figure}[h]
  \includegraphics[width=\columnwidth]{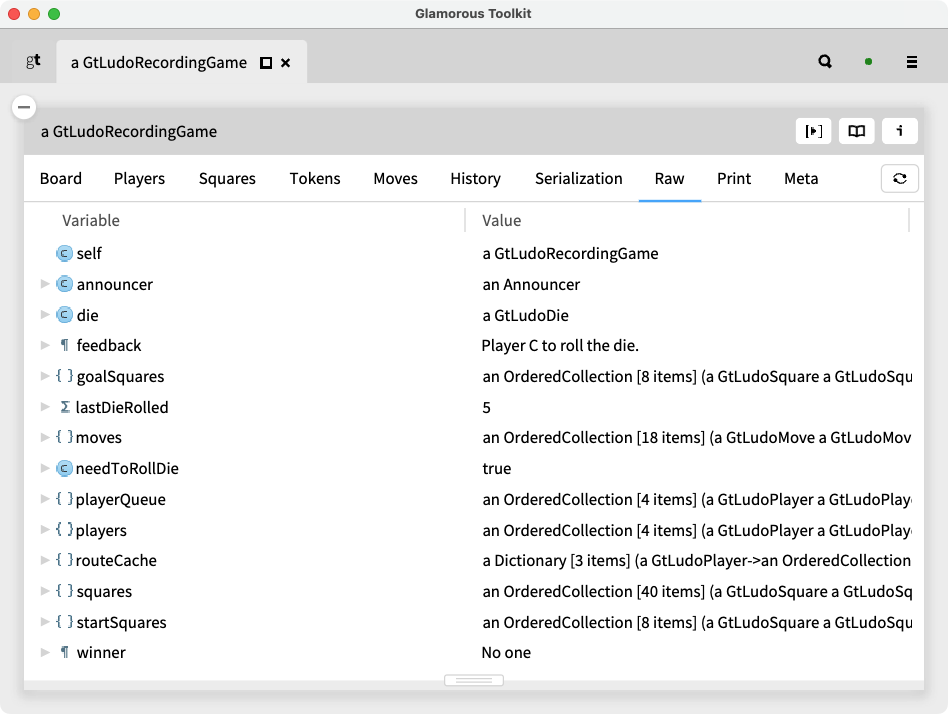}
  \caption{A classical ``raw'' view of a Ludo game.}
  \label{fig:ludoRawView}
\end{figure}

\begin{figure}[h]
  \includegraphics[width=\columnwidth]{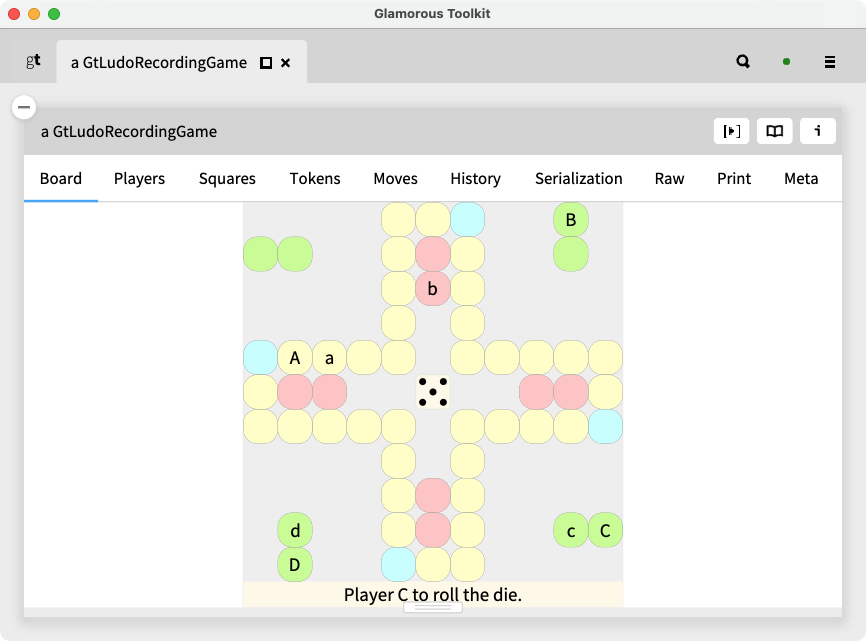}
  \caption{An interactive GUI view of a Ludo game.}
  \label{fig:ludoBoardView}
\end{figure}

Let us consider a typical example.
Suppose we have an implementation of a Ludo\footnote{A simple game in which players move tokens around a board based on the roll of a die.
\url{https://en.wikipedia.org/wiki/Ludo}} game.
Players alternate in throwing a die and moving a token until one player reaches their goal square.
In \autoref{fig:ludoRawView} we see a classical \emph{Raw} view supported by a typical object inspector.
It just shows the state of the object as a list of instance variables and their values.
In \autoref{fig:ludoBoardView}, however, we see an alternative, graphical \emph{Board} view showing the current state of the game as a user would see it.

\begin{figure}[h]
  \includegraphics[width=\columnwidth]{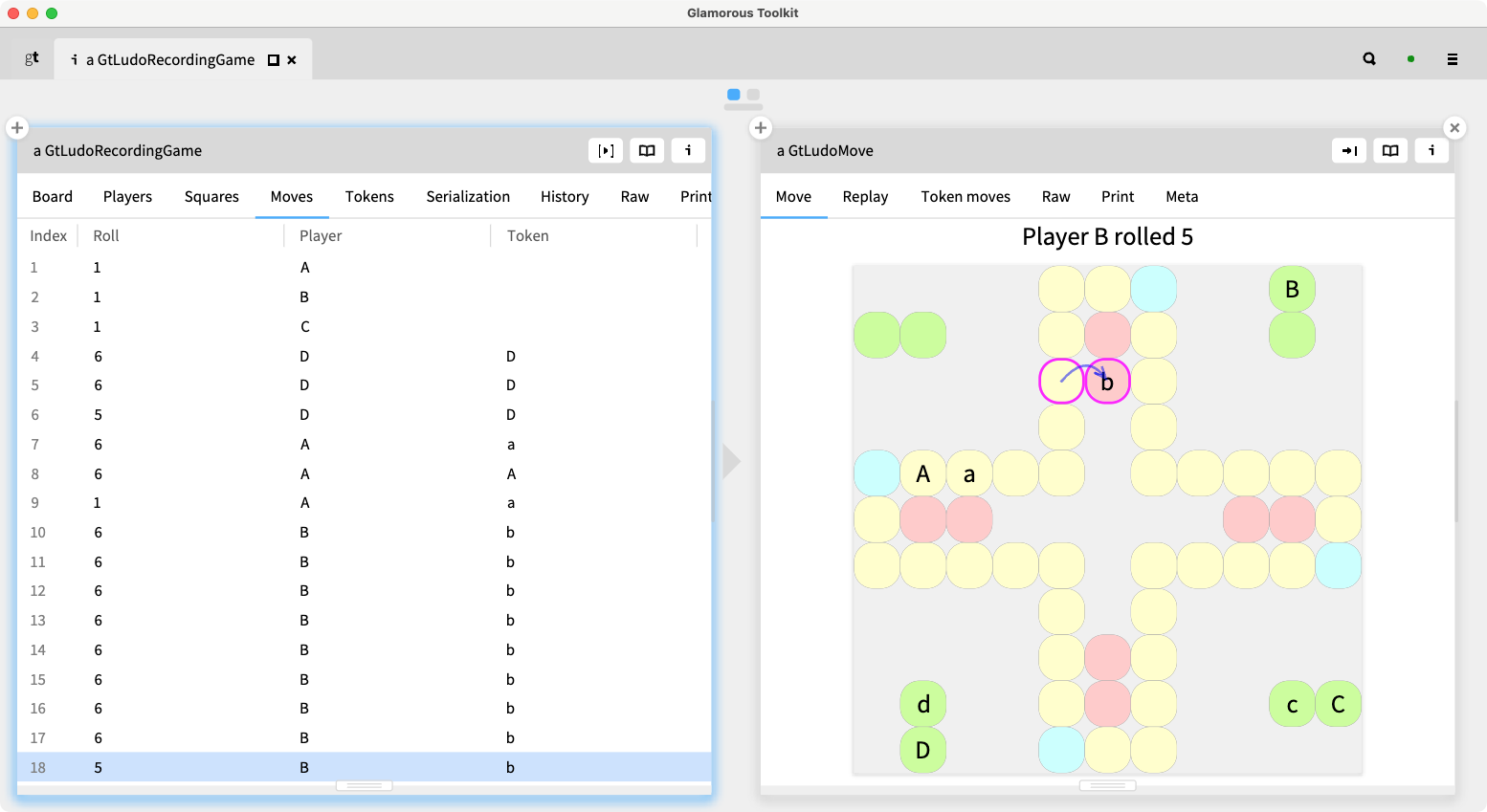}
  \caption{A custom historical \emph{Moves} view of a Ludo game, and a custom graphical view of a selected move.}
  \label{fig:ludoMovesMoveViews}
\end{figure}

In general, a variety of views may be more useful than the classical ``raw'' view, and this is also true for the Ludo game.
In \autoref{fig:ludoMovesMoveViews} we see a \emph{Moves} view that lists all the moves of the game played thus far.
By clicking on a move (at the left) you can then inspect (at the right) the corresponding \st{GtLudoMove} object, with its custom views, in this case showing the details of what happened in this move.

\begin{figure}[h]
  \includegraphics[width=\columnwidth]{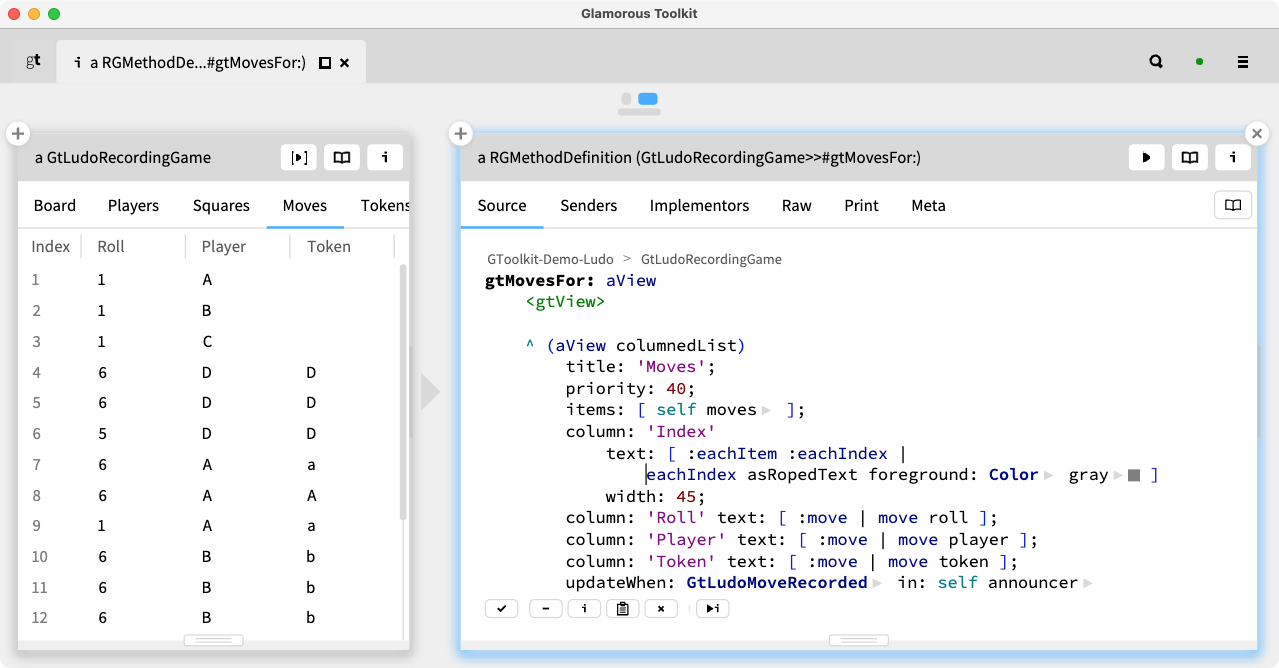}
  \caption{The \emph{Moves} view next to its source code.}
  \label{fig:ludoMovesSource}
\end{figure}

In \autoref{fig:ludoMovesSource} we see the source code defining the \emph{Moves} view.
This method uses just a few lines of code to create a browsable ``columned list'' of past moves with several columns for the details of each move.

In addition to views, objects can also define \emph{actions} that encapsulate useful operations.
For example, the game object in \autoref{fig:ludoBoardView} defines an \lst{Autoplay one move} action (with the icon \smallimg{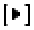}) that can be used to simulate a player's turn.


View and action methods are recognized by the moldable inspector through dedicated annotations, in exactly the same way that a test runner tool in a classical IDE recognizes Java test case methods because they are tagged with a \st{@Test} annotation.
In this case the annotation is \st{<gtView>} for views (and \st{<gtAction>} for actions), seen in the second line of the method defining the view.

\section{Adding Custom Debugger Views}\label{sec:views}

Moldable exceptions provide custom debugger views and actions in essentially the same way as the moldable inspector.
Moldable exceptions are instances of an \emph{Exception} class that has been extended with a dedicated method for each custom debugger view or action.
These methods are annotated with a \lst{<gtExceptionView>} pragma for views and a \lst{<gtExceptionAction>} pragma for actions to extend the moldable debugger
(\ie as opposed to \st{<gtView>} and  \st{<gtAction>} which extend the moldable inspector).

\begin{figure}[h]
  \includegraphics[width=\columnwidth]{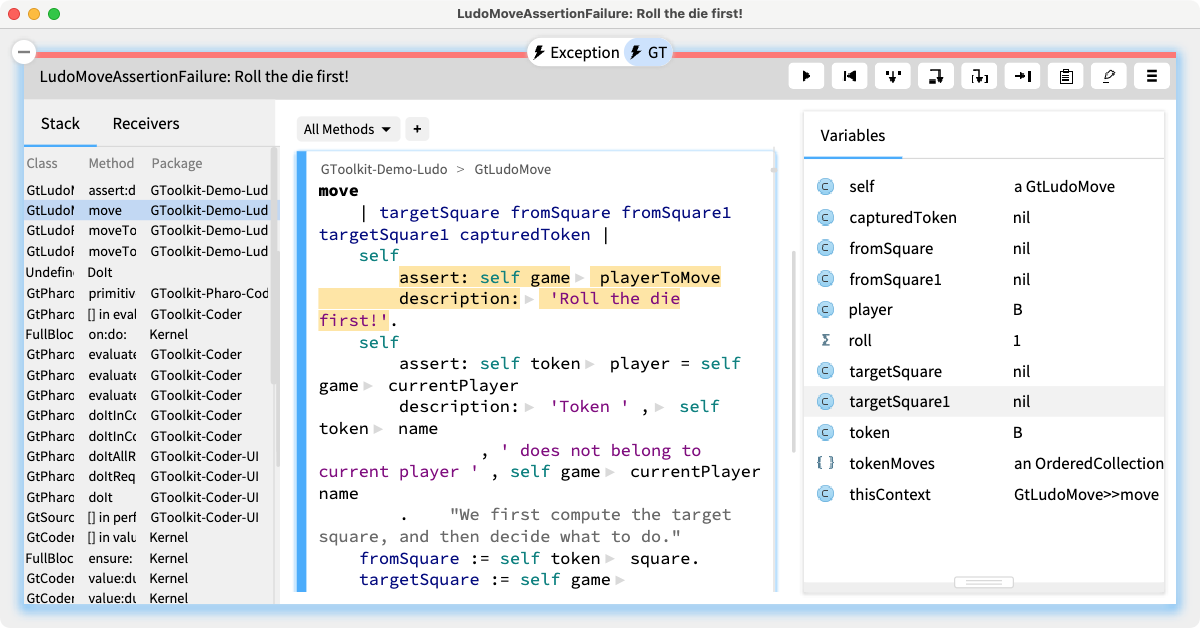}
  \caption{A classical debugger for a precondition failure.}
  \label{fig:ludoClassicDebugger}
\end{figure}

Let us consider that the Ludo game has been implemented with the help of \emph{Design by Contract}~\cite{Meye92b}.
Rolling a die when a player should move, or vice versa, constitutes a precondition violation, which raises a \st{LudoMoveAssertionFailure}.
Similarly, if an attempt is made to move the wrong player's token, this will trigger a precondition failure.
Normally, this would fire up the classical debugger, as seen in \autoref{fig:ludoClassicDebugger}.
Although the precondition violation is clearly reported, the debugger interface is not ideal for tracking down the actual reason for the violation.

What we would perhaps like to see instead is the current state of the game, in addition to a history of the past moves.
We could possibly find these by navigating through the existing debugger views, but why not show them directly?
After all, we know whenever this exception is raised, what it is that we would like to see.
Furthermore, if we already have such views defined elsewhere (we do!), it is not a question of defining new views, but of reusing them in the context of the debugger.
We just need to define two new view methods in the class \lmaf.

Here is the definition of the first view, which simply forwards (delegates) the view to another existing one.
Specifically, the moldable debugger game view for a \lmaf simply reuses the moldable inspector view called \st{gtPositionsFor:} that is already defined in the \st{game} object.
\begin{code}
gtGameViewFor: aView
	<gtExceptionView>
	^ aView forward
		title: 'Game';
		priority: 10;
		object: [ move game ];
		view: #gtPositionsFor:
\end{code}


Let us step through the code:
\begin{inparaenum}[(i)]
	\item \st{gtGameViewFor:} is the name of the view method, which takes as its argument \lst{aView}, the view to be defined.
	\item The method is annotated with \lst{<gtExceptionView>}, which tells the moldable debugger to enable the view whenever \lmaf is raised.
	\item We return (\st{^}) the result of sending \st{forward} to \st{aView}.
	\item We give the view a title and
	\item a priority (the order in which views are shown\,---\,we set it to $10$ rather than $1$ to leave room for more important views later).
	\item We specify the object to forward to (the move's \st{game}).
	\item We name the already existing view method to forward to (\st{gtPositionsFor:}).
\end{inparaenum}

\begin{figure}[h]
  \includegraphics[width=\columnwidth]{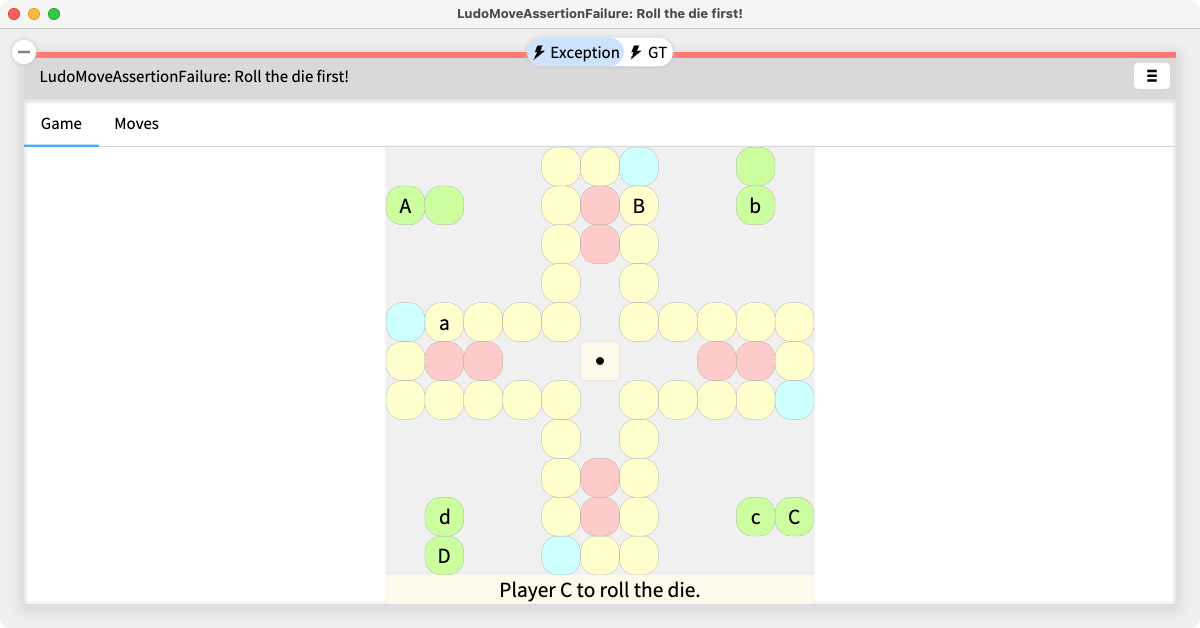}
  \includegraphics[width=\columnwidth]{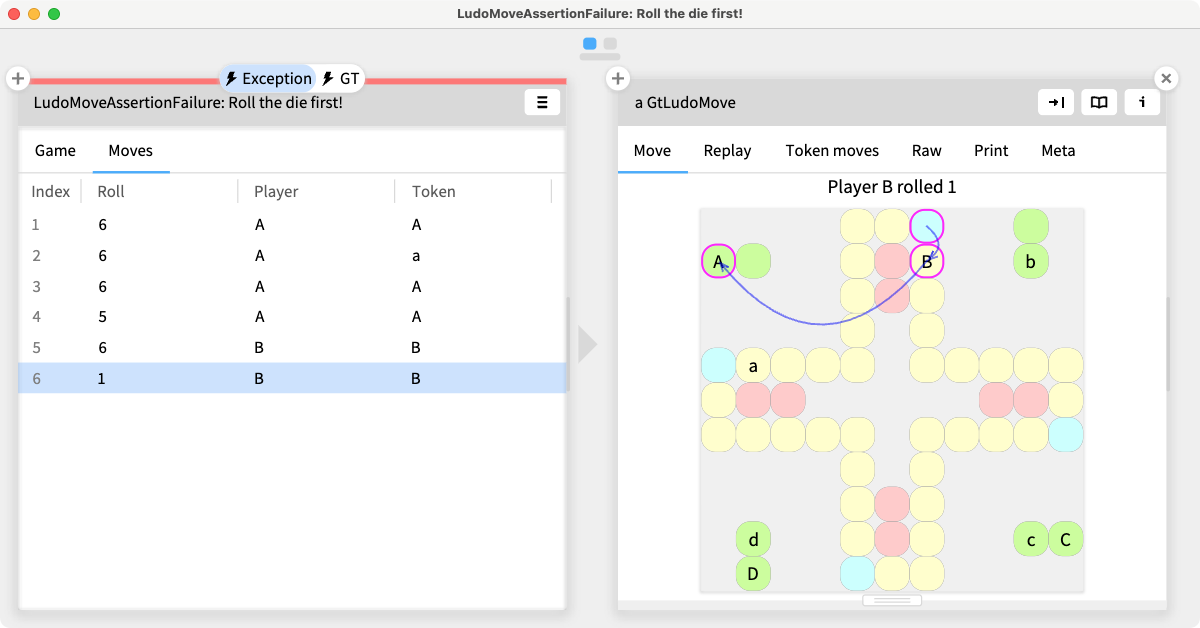}
  \caption{Custom debugger views for the Ludo game.}
  \label{fig:ludoCustomViews}
\end{figure}

We similarly define a method for the history of past moves, and now the debugger, instead of showing us the classical debugger, will offer us the two views seen in \autoref{fig:ludoCustomViews}.
The Game view shows us the current game state graphically, and the Moves view shows us a browsable list of past moves.
Note that we can always switch to the standard debugger by selecting the \emph{GT} button at the top.

The \emph{Moves} debugger view is also defined by reusing the existing object inspector \emph{Moves} view:
\needlines{3}
\begin{code}
gtMovesViewFor: aView
	<gtExceptionView>
	^ aView forward
		title: 'Moves';
		priority: 20;
		object: [ move game ];
		view: #gtMovesFor:
\end{code}

The assertion diff debugger view we saw earlier in \autoref{fig:stringComparisonView} is similarly defined as a method of \st{AssertionFailure}.
\begin{code}
AssertionFailure>>gtComparableTypesTextualDiffFor: aView
	<gtExceptionView>
	| assertionContext |
	self gtHasStack ifFalse: [ ^ aView empty ].
	assertionContext := self gtLocateAssertEqualsContextWithComparableTypes.
	!{\bfseries{assertionContext ifNil:}! [ ^ aView empty ].
	^ aView forward
		title: 'Textual Diff';
		priority: 0;
		object: [ assertionContext ];
		view: #gtComparableTypesTextualDiffFor:
\end{code}
The key difference is that not every \st{AssertionFailure} is raised as the result of a comparison.
For this reason, in line $6$, the view will be suppressed (\st{^ aView empty}) in case the assertion did \emph{not} fail in the context of an \st{assert:equals:} check.

\section{Building Domain-specific Debugger Views}\label{sec:interactions}


The examples we have seen so far have reused existing inspector views, but sometimes there is a need to develop a new kind of debugger view.
This also need not necessarily imply a heavy implementation effort.

The \GT \emph{Scripter} is a tool used to ``script'' GUI interactions, mainly for testing purposes.
Let us consider the case of assertion errors being raised while testing a user interface interaction with the help of the Scripter.
The scenario is the following:
\begin{inparaenum}[(i)]
	\item We create a notebook page called ``Page One''.
	\item We add a text snippet with a link to the \st{Object} class.
	\item We assert the existence of the page.
	\item We click the link.
	\item We check that a new page is created with a code browser for the \st{Object} class.
\end{inparaenum}
In case something goes wrong, we would like the debugger to provide us with high-level views of the state of the Scripter.

\begin{figure}[h]
  \includegraphics[width=\columnwidth]{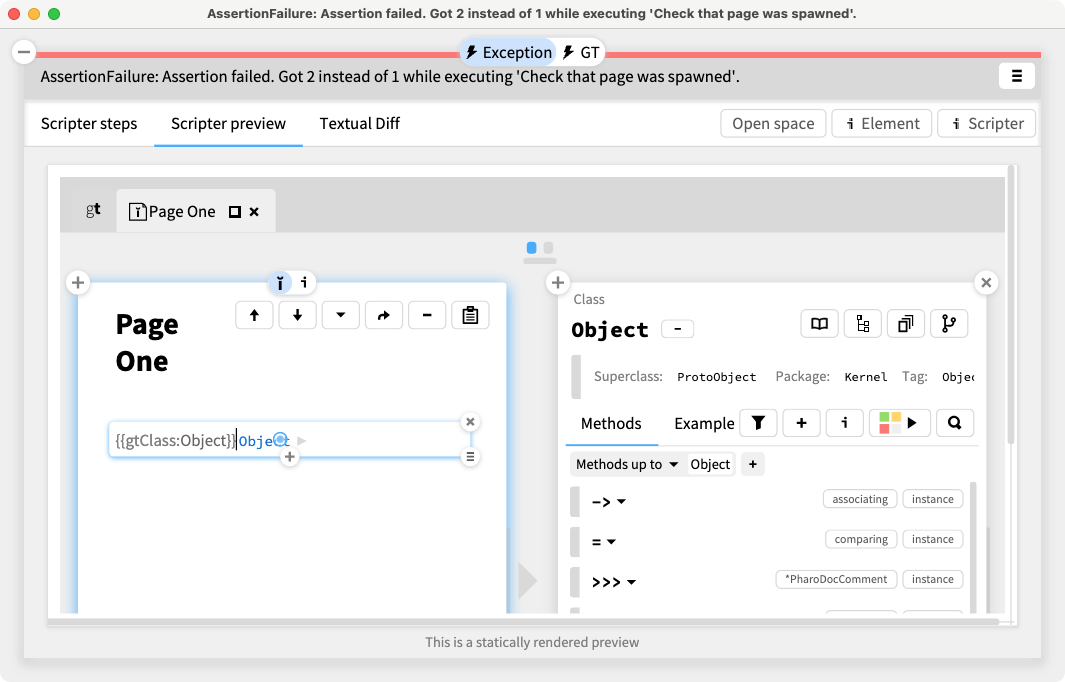}
  \caption{Scripter preview.}
  \label{fig:scripterPreview}
\end{figure}

We consequently introduce three dedicated debugger views.
The \emph{Scripter preview} (\autoref{fig:scripterPreview}) shows the result of the scripted interaction:
\begin{inparaenum}[(i)]
	\item the notebook page has been created, containing a link to the class \st{Object}, and
	\item the link to the class has been clicked, opening a source code browser on the class.
\end{inparaenum}

\begin{figure}[h]
  \includegraphics[width=\columnwidth]{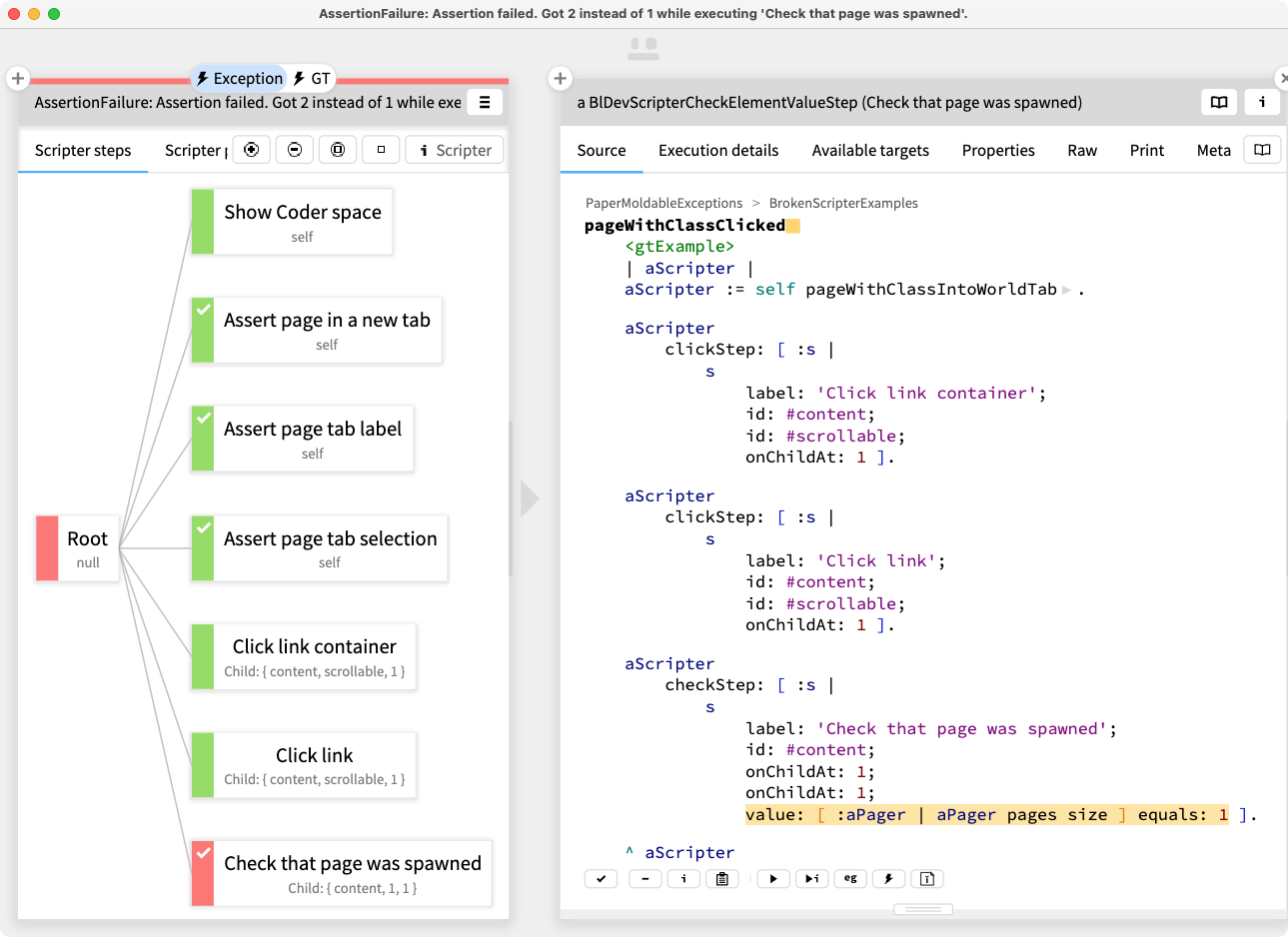}
  \caption{Scripter steps view.}
  \label{fig:scripterStepsViewClicked}
\end{figure}

The \emph{Scripter steps} view (\autoref{fig:scripterStepsViewClicked}) shows us a graphical tree view of the steps performed by the GUI Scripter, as well as the assertions that have been checked.
(In this case the tree has only two levels, but in general there may be many levels.)
The green steps and assertions have succeeded, whereas the red ones have failed.
By clicking on any step or assertion node, we can see the corresponding code highlighted in the scripter method at the right.
Here we see that the step \emph{Check that page was spawned} has failed, and at the right we see the corresponding failure highlighted in the code.

\begin{figure}[h]
  \includegraphics[width=\columnwidth]{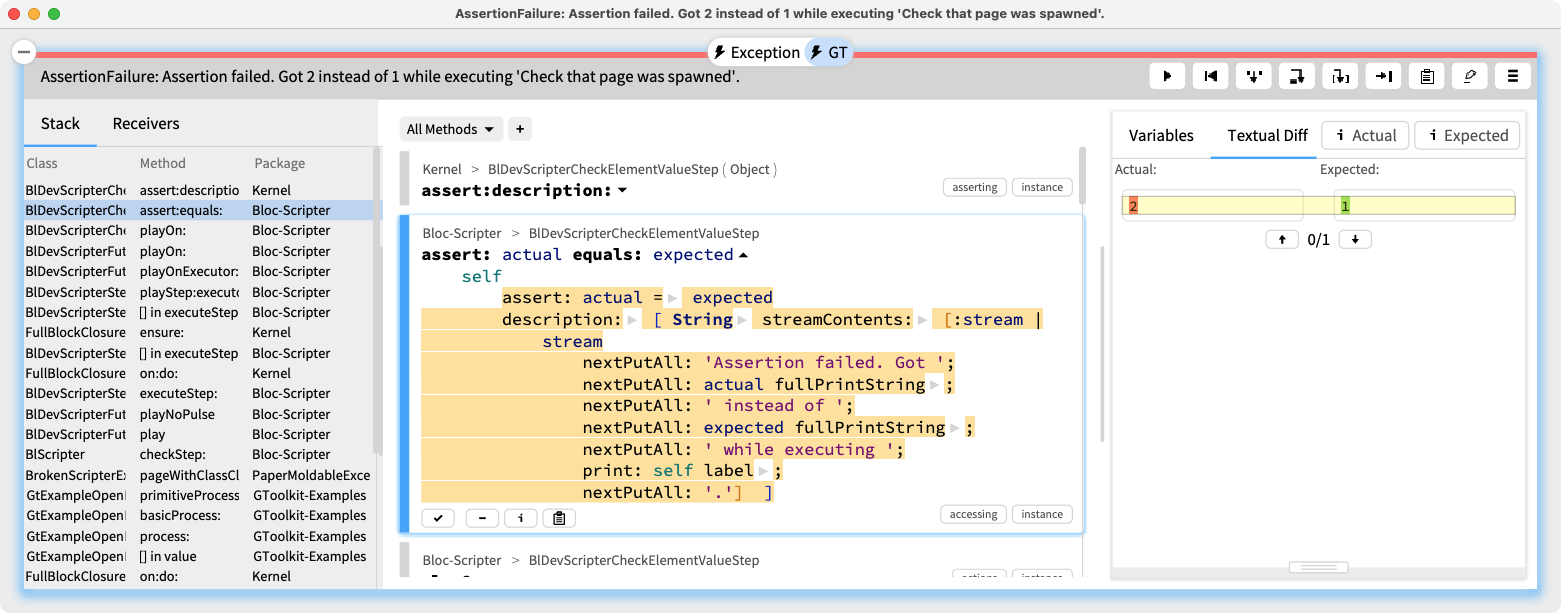}
  \caption{Scripter textual diff view.}
  \label{fig:scripterDiffInDebugger}
\end{figure}

Finally, the \emph{Textual Diff} view (\autoref{fig:scripterDiffInDebugger}) is the same one we have seen earlier, reused again.
It just tells us that the check should verify that there are now two pages in the pager, not just one, as we can also see in the \emph{Scripter preview}.
It is the check specification that is at fault, not the GUI Interaction we are testing.

However this time we show the \lst{Textual Diff} view both in the new debugger interface created for this exception (\lst{Textual Diff} tab in \autoref{fig:scripterPreview}), as well as an additional view within the standard \GT debugger (\lst{Textual Diff} tab in \autoref{fig:scripterDiffInDebugger}).
Hence, a moldable exception can also provide views and actions for the standard \GT debugger.


\paragraph{Why do we need these views?}
The problem with the standard debugger is that, due to the way the GUI Scripter schedules the steps, the offending method (\ie the \lst{page\-With\-Class\-Clicked} method we see in \autoref{fig:scripterStepsViewClicked}) is \emph{not on the stack} at the point where the exception is raised.
Although it is possible to get at the information we seek, doing so is clumsy since we would have to navigate to the exception instance itself, and explore its  internal state, and the classical stack view only confuses matters instead of helping us to debug the problem.
This can be the case with many application domains, especially those that depend on event scheduling.
The run-time stack does not do a good job of telling us how the events have been triggered, so another kind of view is needed.

In general, custom debugger views exist to save you the trouble of rummaging around the stack to find the information you need, and they can offer dedicated custom actions to fix the problem.

\begin{figure}[h]
  \includegraphics[width=\columnwidth]{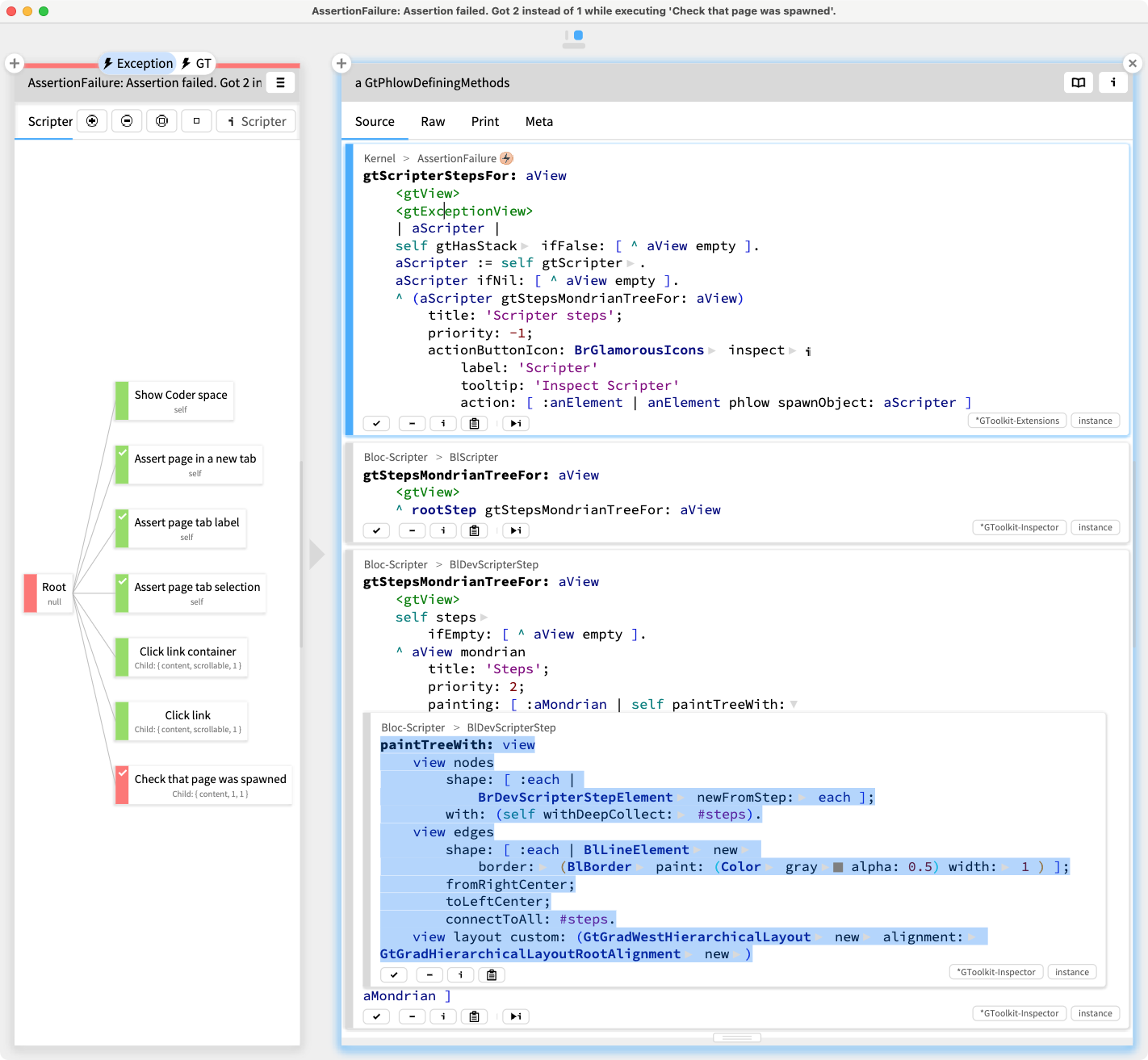}
  \caption{The Scripter steps source code.}
  \label{fig:scripterStepsViewSource}
\end{figure}

\paragraph{How hard is it to implement a domain-specific debugger interface?}
Extending a standard debugger with such views can be challenging.
Typically such views would have to be added in a subclass of the standard debugger.
With moldable exceptions, views only have to be added to the exception class itself.
The \emph{Scripter steps view} is implemented using a version of \emph{Mondrian}~\cite{Pena13b,Meye06a}, a builder for graph-based visualizations.
In \autoref{fig:scripterStepsViewSource} we can see the entire source code of the view expanded in place, implemented in just four methods.\footnote{NB: The figure is intended just to emphasize the size of the source code, not to be an invitation to try to read it.}
The debugging view at the top just adds a ``Inspect Scripter'' button to the next method, an object inspector view for a Scripter.
This in turn just delegates to the \emph{Steps} view of a \st{BlDevScripterStep} object.
Finally, this method embeds a Mondrian visualization implemented in 12 lines of code in the \st{paintTreeWith:} method.

Obviously this does not prove that all domain-specific debugger interfaces will be so tiny, but it does demonstrate that a useful custom debugger view can be implemented in an almost trivial amount of code.

\section{Enabling Automated Fixes}\label{sec:fixes}

\begin{figure}[h]
  \includegraphics[width=\columnwidth]{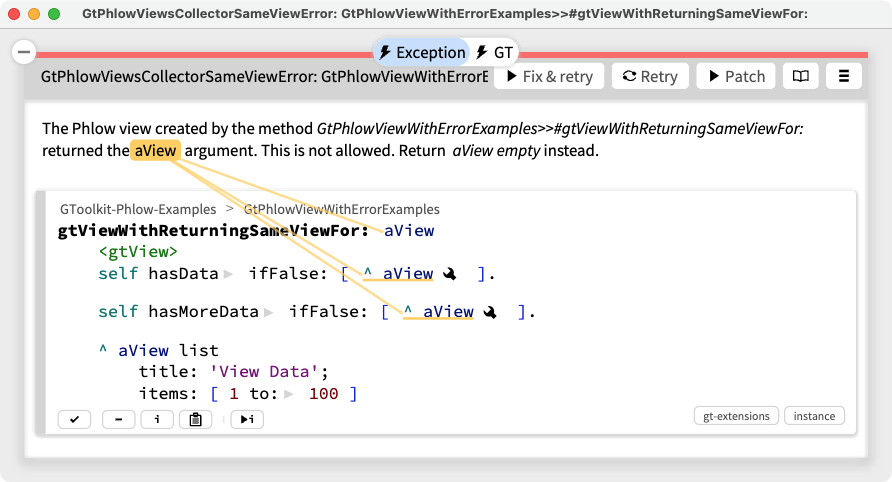}
  \caption{Catching an ``empty view'' error.}
  \label{fig:emptyViewError}
\end{figure}

Many cases of common programming errors can be automatically repaired.
Consider the case
providing to an API an object of the wrong type, or of the right type but the wrong state.
While the first kind of mistake could arguably be caught by a static type system, detecting objects being in a wrong state is rather a run-time issue, typically caught by a precondition.
Some common cases can be fixed by rewriting the client code.
In \autoref{fig:emptyViewError} we see the source code for a custom inspector view that incorrectly returns \st{aView} instead of \st{aView empty} in the preamble.
This custom debugger view decorates the source code with an explanation pointing out the likely errors, and links (in yellow) to the offending source code.

\begin{figure*}[h]
  \includegraphics[width=\textwidth]{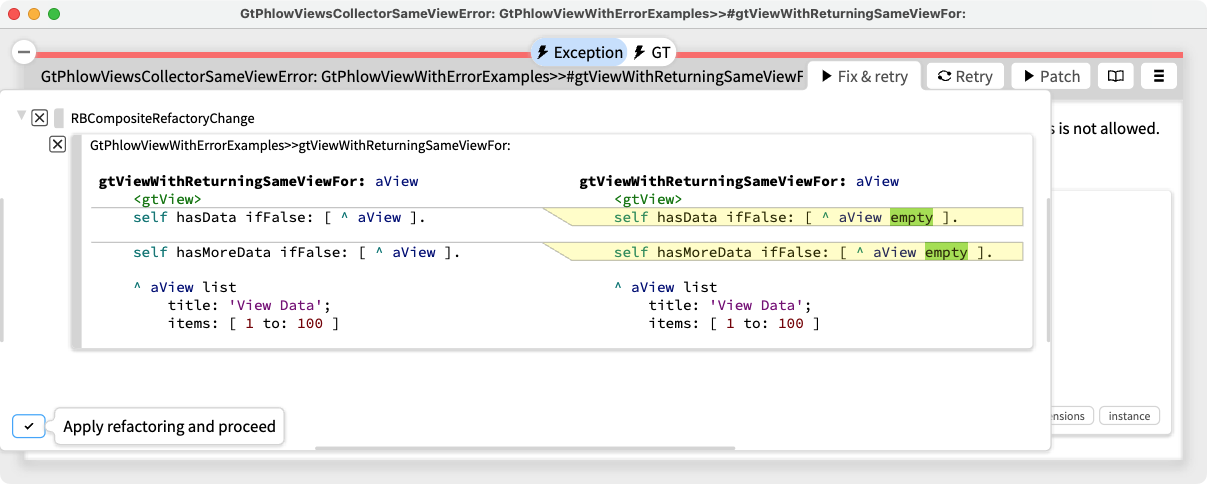}
  \caption{Transforming an empty view error.}
  \label{fig:emptyViewErrorFix}
\end{figure*}

Since such preambles are a common idiom in defining \GT inspector views, and the error is also not uncommon, it becomes easy to fix with the help of a transformation.
\autoref{fig:emptyViewErrorFix} shows the result of performing the \emph{Fix \& retry} action: a refactoring widget is generated that proposes a code transformation.
Clicking on the \emph{Apply refactoring and proceed} button will transform the source code, leave the debugger, and evaluate the rewritten code.
Given how GT inspector views are executed, in this particular case, after applying the transformation that fixes the code of the view, it is enough to restart the top call frame.
In other cases it might be needed to walk over the call stack to find the right place from where to restart execution.

A custom debugger action, such as \emph{Fix \& retry} or \emph{Retry} is defined in the same way as a custom debugger view: it is a method with a particular annotation, in this case being \mbox{\st{<gtExceptionAction>}} rather than \mbox{\st{<gtExceptionView>}}, but  the mechanism is essentially the same. We see the code of this action in the listing below.
Since this is a debugging action we need to be able to control the run-time execution.
To make this possible, methods creating debugging views and actions can take as a second parameter a context object that gives them access to the current run-time execution.


\begin{code}
gtDebugActionFixItFor: anAction inContext: aContext
	<gtExceptionAction>
	(self shouldApplyInContext: aContext )
		ifFalse: [ ^ anAction noAction ].
	^ anAction dropdown
		label: 'Fix & retry';
		icon: BrGlamorousVectorIcons play ;
		priority: 50;
		id: #'gtDebugAction-fixit';
		preferredExtent: 650@350;
		content: [ self refactoringElementWithAction: [
			aContext debuggingSession restartTopContext.
			aContext debugger resumeAndCloseDebugger ]]
\end{code}

The default behavior of such fixits is to open a debugger with the possibility of applying a proposed code transformation.
If, on the other hand, such code transformations should be applied automatically, this can be configured.
In our proof-of-concept implementation of moldable exceptions, this is done by evaluation the following code, which sets a flag in a globally accessible Singleton.
\begin{code}
GtMoldableExceptionTransformationsSettings defaultInstance allowAutomaticTransformations.
\end{code}
The mechanism for managing unhandled exceptions can then optionally consult this flag to decide whether to spawn a debugger or apply the fix (\autoref{sec:implementation} gives more details).

As before, the custom debugger views and actions are mostly built from existing components, so the implementation effort is low.

\section{A Case Study of Moldable Exceptions}\label{sec:casestudy}

As a case study, we applied moldable exceptions within a company that develops applications for managing insurance policies, to improve the development process by integrating custom comparison views into the debugger.

One part of the application consists in generating various documents related to insurance contracts.
The creation of documents is verified using tests that check that a recreated document matches an expected version that has been previously saved.
Whenever a comparison fails, a developer needs to assess the differences between the two documents, and then, in case those differences are justified, update the saved version with the recreated one, or in case they are not expected, track down and fix the root cause of the difference.

In the established workflow, a developer uses a dedicated tool to compare the recreated and the expected versions, and in case the difference is acceptable, update the expected document with the new one.
In case the difference indicates a bug, the developer has to run the test again using the debugger to explore the live execution.
With this workflow, developers have to choose the tool \emph{before} they run the test, which means that they typically need to run a test twice.
This can be a problem as some tests take minutes to run.

With moldable exceptions we integrated custom debugger views that show the differences between the recreated and the expected versions, and provide custom actions to update the expected document with the new one and resume executing the test, directly in the debugger.
To achieve this we rely on the fact that tests raise \lst{ComparisonFailure} exceptions when two documents do not match.

\begin{figure*}[h]
  \includegraphics[width=\textwidth]{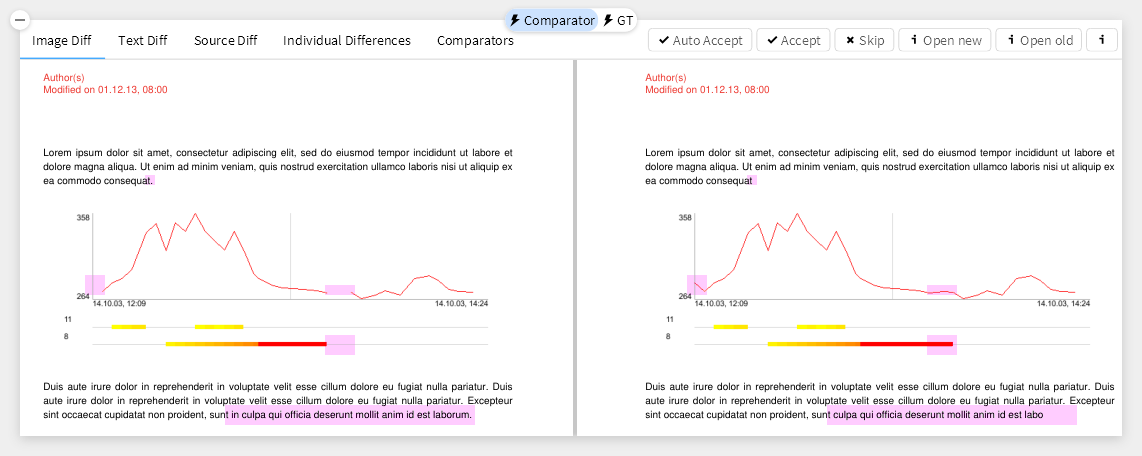}
  \caption{A view showing visual differences between two PDF documents (differences are highlighted in magenta).}
  \label{fig:comparisonFailureVisualDiff}
\end{figure*}

\begin{figure}[h]
  \includegraphics[width=\columnwidth]{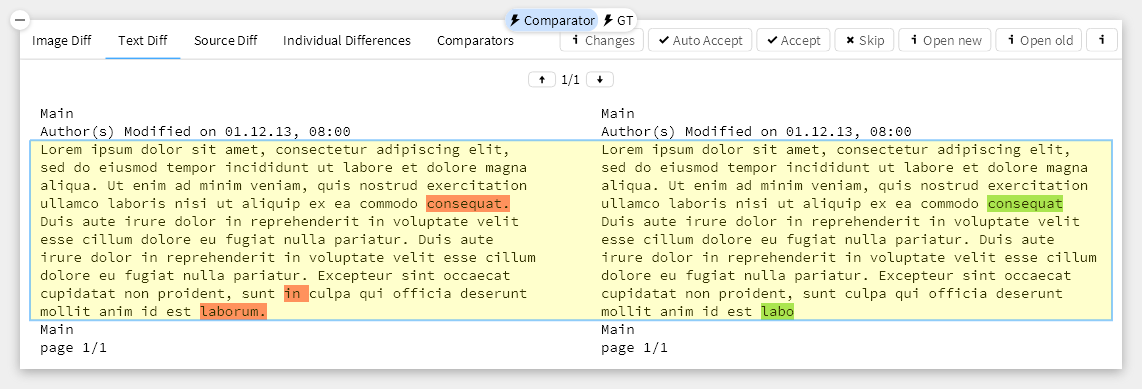}
  \caption{A view showing textual differences between the text content of two PDF documents.}
  \label{fig:comparisonFailureTextualDiff}
\end{figure}

\begin{figure}[h]
  \includegraphics[width=\columnwidth]{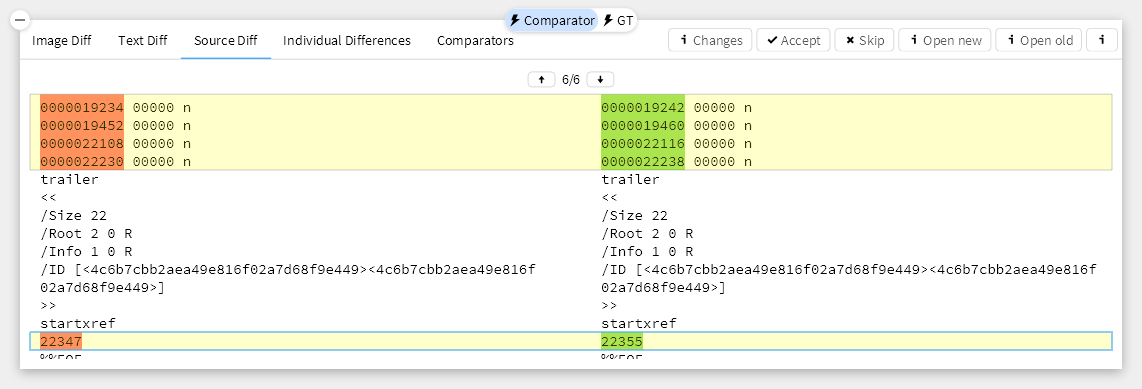}
  \caption{A view showing textual differences between document at the level of the PDF format.}
  \label{fig:comparisonFailureSourceDiff}
\end{figure}

\autoref{fig:comparisonFailureVisualDiff} shows the debugger opened as a result of a failed comparison of two PDF documents.
In this case the exception provides multiple diff views:
one at the image level that visually highlights the parts that differ between the two documents (\autoref{fig:comparisonFailureVisualDiff}),
a second one at the level of the text contained by the PDF documents (\autoref{fig:comparisonFailureTextualDiff}),
and another at the level of the raw document format (\autoref{fig:comparisonFailureSourceDiff}).
Apart from the diff views, the exception also provides an \emph{Accept} action that replaces the saved expected document with the new one, and continues the execution of the test.

The \emph{Image Diff} view and the \emph{Accept} action both reuse existing logic, and in a few lines of code integrate it into the debugger.
This integration unifies the flow of comparing two documents and exploring why the failure happens using a debugger.
Developers can now simply run tests, and when a comparison fails choose how to proceed.

%
%

\section{Implementation Details}\label{sec:implementation}



The key to implementing moldable exceptions is the mechanism used to inform a moldable debugger of the relevant views and actions it provides.
In our implementation every exception can provide one or more \emph{debugger specifications}.
Exceptions define these specifications through methods annotated with \lst{<gtDebuggerSpecification>}.
Below we see the method in the \lst{Exception} class, at the top of the exception class hierarchy, that defines a custom debugger for every exception.

\begin{code}
Exception>>gtExceptionDebuggerSpecification
	<gtDebuggerSpecification>
	^ GtMoldableExceptionSpecificDebuggerSpecification
		forException: self
\end{code}

\begin{figure}[h]
  \includegraphics[width=\columnwidth]{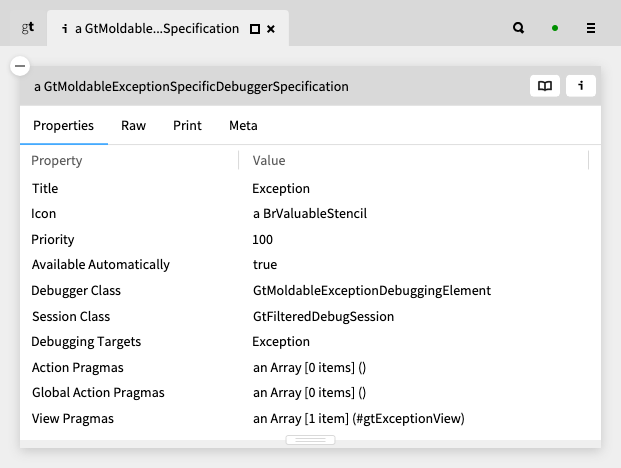}
  \caption{Default properties for a debugging specification.}
  \label{fig:debuggerSpecificationsProperties}
\end{figure}

A debugger specification contains all the information needed by a debugger to locate actions and views.
\autoref{fig:debuggerSpecificationsProperties} shows the default properties defined by such a specification.

By default, as mentioned in \autoref{sec:views}, the moldable \GT debugger looks for methods annotated with a \lst{<gtExceptionView>} pragma.
An exception can change the pragma if needed, or provide a list of other pragmas.
The default specification also provides several other properties, including:
\begin{itemize}
\item Debugging Targets: objects reachable from the exception where to look for actions and views.
\item Debugging Actions: pragmas for debugging actions.
\item Title, Icon: for displaying a debugger selector.
\item Priority: a number used to determine the order in which alternative debuggers are offered.
\item Activation Predicate: a condition to decide whether to show the debugger or not based on the exception state; by default the debugger is shown if any debugging views are found.
\item Available Automatically: if the debugger is active, indicates whether it can be shown by default.
\item Debugger Class: The class of the debugger that is created from this specification.
\end{itemize}

By default the debugger created from a specification searches for debugging views and actions in the class of the exception.
However, through the \lst{Debugging Targets} property an exception can change that, and provide any other list of objects.
That makes it possible for other objects in the system reachable from the exception objects to provide debugging actions and views.

\begin{figure}[h]
  \includegraphics[width=\columnwidth]{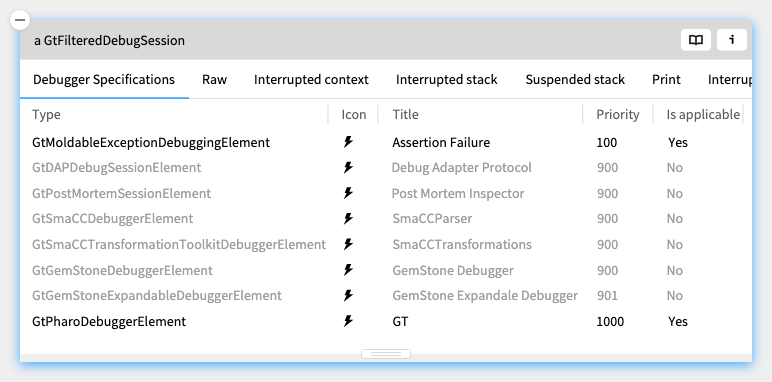}
  \caption{A list of debugger specifications found when an exception is raised.}
  \label{fig:availableDebuggerSpecifications}
\end{figure}

To summarize, when an exception is raised, the moldable debugger looks for debugger specifications in exception methods annotated with  \lst{<gtDebuggerSpecification>} and uses them to create custom debuggers.
These custom debuggers are then shown when their activation predicate is true.
For example, \autoref{fig:availableDebuggerSpecifications} shows the list of debugger specifications located when the assertion failure from \autoref{fig:stringComparisonView} is raised.
We see that in total eight specifications are found, but just the one for the default debugger together with the one provided by the exception are active, and result in a custom debugger view being shown.

To enable moldable exceptions to perform automated transformations instead of opening the debugger, we provide a dedicated trait~\cite{Duca06b} that exceptions can use.
This trait provides an implementation for the signal method shown below.
This method is executed whenever an exception is raised.
By default in case an exception is not handled in the system a debugger is opened.
This override checks if automatic transformations are allowed for the current exception (line 4), and if they are, asks the exception for its transformation (lines 5) and executes it if there is anything to transform (lines 6-7).
Transformations can apply code changes, modify the state of objects, or perform other operations, such as deleting no longer needed files.

\begin{code}
TGtMoldableExceptionSignalWithTransformation>>signal
	<debuggerCompleteToSender>
	| transformation |
	self canApplyAutomaticTransformation ifFalse: [ ^ super signal ].
	transformation := self gtMoldableExceptionTransformation.
	transformation shouldTransform ifFalse: [ ^ super signal ].
	transformation performTransformation
\end{code}


\section{Discussion and Future Directions}\label{sec:directions}

Moldable exceptions are just objects augmented with annotated methods that create alternative views and actions when they are used to spawn a moldable debugger.
The only changes needed to an application are:
\begin{inparaenum}[(i)]
	\item if there is not already a dedicated exception class for the exception to be molded, introduce one, and
	\item add dedicated exception view methods to that class to mold the debugger.
\end{inparaenum}

\begin{figure}[h]
  \includegraphics[width=\columnwidth]{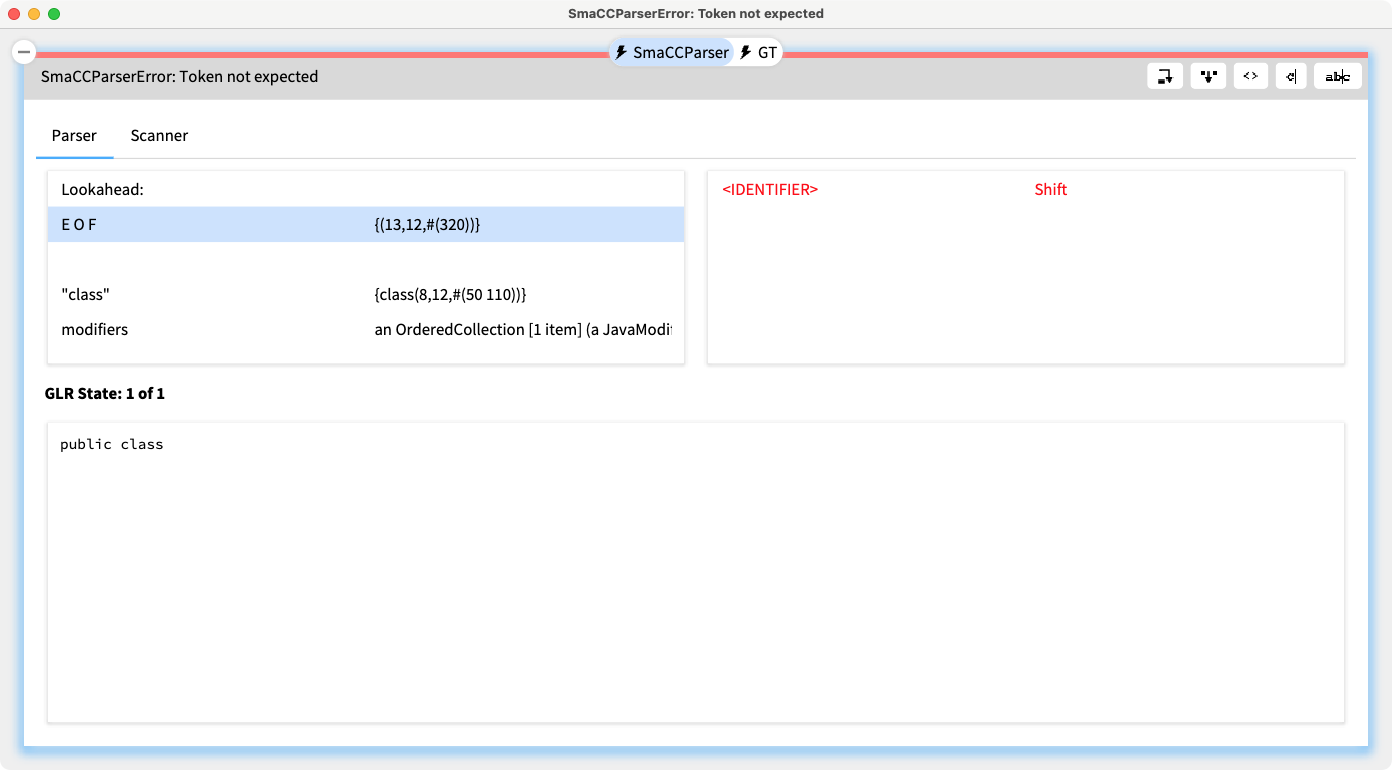}
  \caption{A custom SmaCC debugger on an invalid Java snippet.}
  \label{fig:smaccDebugger}
\end{figure}

The previous extension mechanism for the moldable debugger required developers to extend the debugger by creating subclasses~\cite{Chis15c}.
Although the old framework considerably reduced the cost of building a custom debugger, they still required a non-trivial implementation effort, consisting of several hundreds of lines of code (\cite{Chis15c}, table 7).
While both the moldable inspector and the (old) moldable debugger were available in \GT up to now, currently \GT comes with 3200 views and 280 actions with an average of 12 lines of code, but only 8 custom debuggers (such as the one in  \autoref{fig:smaccDebugger}).


This indicates that the extension mechanism plays a significant role in developers extending the IDE, and that the cost of creating views and actions can be very low.
Due to this, through moldable exceptions we aim to apply the same extension mechanism present in the moldable inspector for creating custom debuggers.
In the new approach, a moldable debugger simply needs to recognize whether a raised exception has custom views defined (in our case by detected annotated methods of its class), and then activate those views in its user interface.


At the time of writing we have implemented some 26 custom debugger views, in an average of under 12 lines of code.
Over half of these are forwarding views that reuse (delegate to) existing views previously defined as object inspector views.
Several more are simple text, list or tree views, and only three create custom graphical widgets.

The simplest debugging views are just like inspector views, and many of our examples just forward (delegate) to existing views, but more generally exceptions have access to the reified run-time stack at the time that they are raised, so debugger views can extract and present arbitrary run-time information.
The \emph{Scripter steps} view we saw earlier in \autoref{fig:scripterStepsViewClicked} offers an example.
The same approach could be used, for example, to present or highlight just the ``interesting'' stack frames, for example in an event-driven application, just the methods that are responsible for processing events.

At this time, moldable exceptions only offer the possibility to provide alternative views and actions, but not alternative ways to step through the execution.
Currently this is only possible in our setting by providing a completely separate debugger implementation.
In \autoref{fig:smaccDebugger} we see such a dedicated debugger for SmaCC~\cite{Brant17a}, the Smalltalk compiler compiler framework that has been spawned on an invalid fragment of Java code.
The SmaCC debugger offers the possibility to step through the grammar rules of a parser and explore its execution state.


Our proof-of-concept implementation of moldable exceptions presented here heavily leverages the existing infrastructure for moldable inspector views in \GT, but in principle there is nothing to prevent its application to other programming languages and IDEs.
The basic idea is simple: any moldable tool must be prepared, whenever it is created, to examine the execution context of the objects it is initialized with, and use that context to adapt its behavior.
In the case of a moldable debugger, this context is provided by the exception raised.
Custom debugger views and actions must then be provided by the specific exception raised.
One way to provide such views and actions is through specially annotated methods, but other means could be used, such as naming conventions, or a registry of debugger extensions.
The precise extension method will depend on the language technology available.


\section{Related Work}\label{sec:related}

Early debuggers provided a command-line interface to inspect and step through the execution of a running program or a \emph{post mortem} ``core dump.''
With the advent of GUI-based IDEs, debuggers were also updated to offer interactive views to explore the execution state of a program and to step through its execution~\cite{Rose96a}.

Studies of developer debugging behavior have shown that many developers have difficulty using debuggers and often shy away from them.
McCauley \etal report:
``We note, however, that our experiences with debugging tools is that many of them use execution traces as a method of assisting students to understand the execution of a program.
Tubaishat~\cite{Tuba01a} characterized the use of execution traces as an example of a shallow reasoning technique''~\cite{McCa08a}.
And Beller \etal note that
``Debuggers are difficult to use.
Another reason given by interviewees, even though seasoned engineers, was that `the debugger is a complicated beast' (I2) and that `debuggers that are available now are certainly not friendly tools and they don't lend toward self-exploration'.''
\cite{Bell18a}

An early example of an extensible debugger is \deet~\cite{Hans97a}.
Extensions to \deet were written Tcl/Tk or the Korn shell.
The focus of the extension mechanism in \deet is on adding new debugger features, rather than providing debugging support for specific application contexts.

The Java Virtual Machine Tools Interface (JVMTI)~\cite{JVMTI24} is an API for tools to inspect the state and control the execution of Java applications.
Agents can be written in any native language that supports C calling conventions.
Custom views must be built from scratch and integrated in your environment.

The VSCode Debugger Adapter Protocol (DAP)~\cite{DAP21} makes it possible to implement a generic debugger for a development tool that can communicate with different debuggers via Debug Adapters.

mbeddr~\cite{Voel17a} is a collection of languages and language extensions, mainly focusing on C, built with the JetBrains MPS language workbench~\cite{Camp14a}.
mbeddr includes an extensible debugger framework~\cite{Pavl15a} that focuses on providing support for the individual language extensions, not application specific customizations.

Bousse \etal have developed a generic omniscient (``back in time'') debugger that can be adapted to different domain-specific languages (DSLs) with moderate effort~\cite{Bous18a}.
They use DAP to provide new controls for DSLs in VSCode~\cite{Enet23a}.
Their work focuses more on the stepping semantics than on providing custom views.

There have been numerous efforts to develop specialized debuggers for DSLs using a variety of approaches.
Wu \etal report on grammar-driven generation of DSL debuggers~\cite{HuiW08a}.
Lindeman \etal leverage the Spoofax language workbench~\cite{Kats10a} to declaratively define DSL debuggers~\cite{Lind11a}.
D2X~\cite{Brah23a} provides an API for DSLs to define debugger extensions.
All of these DSL debugger approaches focus on providing general debugging for a given DSL, not for more finely-grained application debugging issues.

Babylonian programming~\cite{Rau19a} is an approach in which live examples are integrated into the code editor, and offers various specialized  views~\cite{Rein24a} of the run-time behavior of the examples.
The current implementation of Babylonian/S does not offer a means to define application-specific views for examples.

Sindarin is an extensible Pharo-based debugger that offers a rich API for scripting debugger extensions~\cite{Dupr19a}.
The use case for Sindarin is rather different from that of moldable exceptions, allowing developers to enter debugger scripts during a debugging session, rather than as a way to enable specialized debugger behavior in response to specific kinds of exceptions.
It would be interesting to explore the use of Sindarin to implement custom stepping behavior.

The Moldable Debugger~\cite{Chis15c} is an example of a moldable tool~\cite{Chis17a} that adapts its behavior to a specific run-time application context.
The original moldable debugger, however, still required some significant implementation effort to define an alternative debugger interface, much like the other approaches we have cited from literature.
Similarly, the original moldable inspector~\cite{Chis15a} did not offer a very easy mechanism to provide custom inspector views depending on the object being inspected.
The adoption of the extension mechanism of using annotated methods to define inspector customizations led, over time, to the development of a large number of custom inspector views in the current version of \GT (over 3000 at last count).

Disciplined use of exceptions is a well-established practice in object-oriented software development~\cite{Meye88a}.
The key insight and original contribution of moldable exceptions is to leverage exceptions as the hook to molding the behavior of the debugger.
Although it is common practice to add textual descriptions to exceptions, for example, by overriding the \st{toString()} method in Java, associating debugging views to exceptions is novel.
No previous approach to extensible debuggers, including the original moldable debugger, has taken this approach.

\section{Conclusion}\label{sec:conclusion}



Interactive debuggers are all basically the same.
Command-line debuggers offer commands to sample the current execution state and step through the running code.
Graphical debuggers that show us the run-time stack, and offer buttons instead of commands to step through the code, but they are still all the same.
The trouble with this is that every debugging problem is different, but debuggers all show us the same thing, and they offer the same kinds of actions to step through the code.

Our key contribution is to show how a debugger can be dynamically molded with custom views and actions that depend on the specific exception raised.
The mechanism is lightweight, and many useful extensions can be defined with minimal implementation effort.

\begin{acks}
Many thanks to Alexandre Bergel, Steven Costiou, and Timo Kehrer for their reviews of a draft of this paper.
\end{acks}

\bibliographystyle{ACM-Reference-Format}
\bibliography{moldableExceptions}

\end{document}